\documentclass[aps,pra,onecolumn,showpacs,amsmath,amsfonts,amssymb,superscriptaddress]{revtex4}
\usepackage{xspace,epsfig,amsfonts,amssymb}

\begin{document}

\newcommand{\be}{\begin{equation}}
\newcommand{\ee}{\end{equation}}
\newcommand{\ba}{\begin{eqnarray}}
\newcommand{\ea}{\end{eqnarray}}
\newcommand{\la}{\langle}
\newcommand{\ra}{\rangle}
\newcommand{\h}{\hskip 1cm}
\newcommand{\hh}{\hskip 2cm}
\newcommand{\lo}{\longrightarrow}

\title{Photon losses depending on polarization mixedness}

\author{Laleh Memarzadeh}
\email{laleh.memarzadeh@unicam.it}

\affiliation{Dipartimento di Fisica, Universit\`{a} di Camerino,
I-62032 Camerino, Italy}

\author{Stefano Mancini}
\email{stefano.mancini@unicam.it}

\affiliation{Dipartimento di Fisica, Universit\`{a} di Camerino,
I-62032 Camerino, Italy}

\date{\today}

\begin{abstract}
We introduce a quantum channel describing photon losses depending
on the degree of polarization mixedness. This can be regarded as
a model of quantum channel with correlated errors between
discrete and continuous degrees of freedom. We consider classical
information over a continuous alphabet encoded on weak coherent
states as well as classical information over a discrete alphabet
encoded on single photons using dual rail representation. In both
cases we study the one-shot capacity of the channel and its
behaviour in terms of correlation between losses and polarization
mixedness.
\end{abstract}


\maketitle


\section{Introduction}

In recent years lots of attention has been devoted to capacity of
quantum channels and related properties \cite{KingWinterHasting}.
Along this direction introducing models with correlated noise has
provided more realistic framework for studying quantum channels
\cite{KW}. Correlated noise was first introduced in \cite{M2} for
depolarizing channel, then in subsequent works \cite{M3, disMar}
the advantage of using entangled states was addressed for
different channels and some of the effects has also been
experimentally demonstrated \cite{Banaszek}. The notion of
quantum channels with correlated noise has also been extended to
continuous-variable, e.g. lossy bosonic channels \cite{lbmc} and
channels with additive noise \cite{M4}.

In all the earlier works, correlation is defined between
different uses of the channel, but within the same degree of
freedom. In practice correlated errors may even arise in a single
channel use between different degrees of freedom. Optical fibers,
which are among the most successful candidates for quantum
communication, are examples of this kind. In any use of an
optical fiber, different types of noise may happen like loosing
photons to the environment and at the same time having
polarization effects. Loss effects in optical fibers can be
modeled by a beam splitter interaction with a vacuum (or thermal)
environment. The classical capacity of the memoryless lossy
bosonic channel was exactly evaluated in \cite{GioPRL92}, while
for the same channel with memory the evaluation turns out to be
more involved and to be model dependent \cite{lbmc}.

Although the capacity of lossy bosonic channels has been widely
investigated, the study of polarization effects in the rate of
information transmission is more restricted to classical optical
communication \cite{Gisin95}. In this paper we introduce a
quantum channel describing photon losses depending on the degree
of polarization mixedness. This can be regarded as a model of
quantum channel with correlated errors between different degrees
of freedom. We consider classical information over a continuous
alphabet encoded on weak coherent states as well as classical
information over a discrete alphabet encoded on single photons
using dual rail representation. In both cases we study the
one-shot capacity of the channel and its behaviour in terms of
correlation between losses and polarization mixedness.

The layout of the paper is the following. In section \ref{Model}
we introduce the model.
 We then study  in section \ref{wcInput} the behaviour of Holevo information
versus the correlation parameter for the case of information
encoded in weak coherent states. In contrast, in section
\ref{pInput} we use the dual-rail encoding for qubits. In this
case our model realizes a quantum erasure channel \cite{erasure}
where the degree of correlation controls the probability of
erasing information. Paper ends with conclusion in section
\ref{conclusion}.


\section{The Model}\label{Model}

\begin{figure}[t]
 \centering
  \includegraphics[width=4.5cm,height=5cm,angle=0]{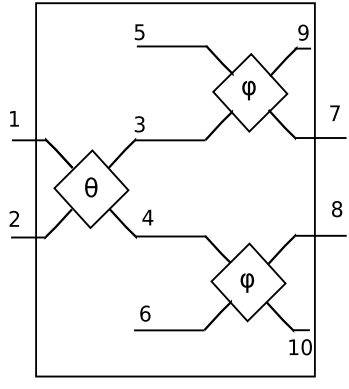}
 \caption{Polarization mixing and losses are modeled by three beam splitters characterized by real random variables $\theta$ and $\phi$
 which have the probability distribution $P(\theta,\phi)$.
 Input states enter the channel in modes $1$ and $2$ and the output is read in modes $7$ and $8$.}
\label{channel}
\end{figure}

Let us consider a laser pulse which first passes through a
polarization beam splitter to separate two polarizations, e.g.
horizontal and vertical, and then is sent through a fiber. Due to
imperfections on the fiber, the two polarizations do not necessarily
remain distinct, but they tend to become mixed. At the same time
along the travel on the fiber the pulse tends to be attenuated.
This attenuation effect can be related to the polarization mixing
effect. The situation can be modeled by three beam splitters as
shown in figure (\ref{channel}).

In general the evolution operator of a beam splitter is given by
the unitary transform
\begin{equation}
U(\zeta)=e^{\zeta ab^{\dag}-\zeta^*a^{\dag}b},
\end{equation}
depending on a complex parameter $\zeta$. Here
$a$ ($a^{\dag}$) and $b$ ($b^{\dag}$) are annihilation (creation)
operators for the two input radiation modes. It is easy to show
that the Heisenberg evolution of the creation operators of the
two modes of the input state is given by
\begin{equation}\label{U}
U(\zeta)\left(\begin{array}{c} a^{\dag}\cr
b^{\dag}\cr\end{array}\right)U^{\dag}(\zeta)
=\left(\begin{array}{cc} 
\cos|\zeta|& e^{i\arg\zeta}\sin|\zeta| \cr
-e^{-i\arg\zeta}\sin|\zeta| &\cos|\zeta|
\end{array}\right)\left(\begin{array}{c} a^{\dag}\cr b^{\dag}\cr\end{array}\right).
\end{equation}
All the beam splitters in the channel are characterized by real
parameters $\theta$ and $\phi$. Labeling the creation operators of
the input modes by $a_1^{\dag}$ and $a_2^{\dag}$ and using
equation (\ref{U}), the creation operators of modes 3 and 4 in
figure are given by
\begin{eqnarray}
a_3^{\dag}&=&\cos\theta a_1^{\dag}+\sin \theta a_2^{\dag}\cr
a_4^{\dag}&=&-\sin \theta a_1^{\dag}+\cos\theta a_2^{\dag},
\end{eqnarray}
which shows that they are a mixture of modes 1 and 2, hence the
polarization mixing effect. For example, if the mode 1 (2)
denotes the horizontal-polarized (vertical-polarized) photons, in
mode 3 (4) photons with both horizontal and vertical polarization
will appear.

Then, each output of the first beam splitter goes through another
beam splitter, characterized by a real parameter $\phi$,
 and interacts with environment in the vacuum state.
This part models the losses in the channel with a loss rate
equal to $\sin^2\phi$.

The total action of these three beam splitters on the input state
$\rho_{1,2}$ is described by the map
\begin{equation}
\Phi_{\theta,\phi}(\rho_{1,2})={\rm Tr}_{env}\left[
U_{4,6}(\phi)U_{3,5}(\phi)U_{1,2}(\theta) (\rho_{1,2}\otimes
|00\ra_{3,4}\la
00|)U^{\dag}_{1,2}(\theta)U^{\dag}_{3,5}(\phi)U^{\dag}_{4,6}(\phi)\right],
\end{equation}
where the environment modes are labeled by indices $5$ and $6$.
To complete the definition of the channel we introduce
correlation between polarization mixing (characterized by
$\theta$) and losses (characterized by $\phi$). This can be done
by means of a joint probability distribution $P(\theta,\phi)$ for
the values that $\theta$ and $\phi$ can take. Thus, the output of
the channel results
\begin{equation}\label{finalOUT}
\Phi(\rho)=\int d\theta d\phi P(\theta,\phi)
\Phi_{\theta,\phi}(\rho).
\end{equation}
We assume $P(\theta,\phi$) to be a Gaussian distribution of the
following form:
\begin{equation}\label{joint}
P(\theta,\phi)=\frac{1}{2\pi\sigma^2\sqrt{|\gamma^{-1}}|}e^{-\frac{1}{2\sigma^2}
r^t \gamma r },
\end{equation}
where $r=(\theta-\theta^*, \phi-\phi^*)^t$ and $\gamma$ is the
correlation matrix:
\begin{equation}
\gamma=\left(\begin{array}{cc} 1&-x\cr -x&1\end{array}\right).
\end{equation}
The parameter $x\in[0,1]$ accounts for the degree of correlation
between the two effects (polarization mixing and loss). For
$x=0$ we have
\begin{equation}\label{uncorr}
P(\theta,\phi)=p(\theta)p(\phi)
\end{equation}
where $p(\theta)$ and $p(\phi)$ are independent and identical
probability distributions
\begin{equation}
p(\theta)=\frac{1}{\sqrt{2\pi\sigma^2}}e^{-\frac{(\theta-\theta^*)^2}{2\sigma^2}},\hskip
1cm
p(\phi)=\frac{1}{\sqrt{2\pi\sigma^2}}e^{-\frac{(\phi-\phi^*)^2}{2\sigma^2}}.
\end{equation}
They are characterized by mean values $\theta^*$ and $\phi^*$ and
variance $\sigma^2$. By increasing $x$, the distance between the
probability distributions $P(\theta,\phi)$ and $p(\theta)p(\phi)$
increases,
\begin{equation}
\left| P(\theta,\phi)-p(\theta)p(\phi)\right|=p(\theta)p(\phi)\left|
1-e^{-\frac{1}{2\sigma^2}(\theta-\theta^*)(\phi-\phi^*)x}\right|.
\end{equation}
This can be interpreted as increasing the correlation \cite{KMM}.
Since at $x=1$ the maximum of the joint probability distribution
is at $ \theta-\theta^*=\phi-\phi^*$ we consider it as the most
correlated
case.

The last remark which is necessary on the probability
distribution (\ref{joint}) is that it is not a periodic function
of $\theta$ and $\phi$, while the beam splitter operators are
periodic in $\theta\in[-\pi,\pi]$ and $\phi\in[-\pi,\pi]$.
Hereafter we assume the variance $\sigma$ of the probability
distribution to be sufficiently small ($\ll2\pi$), so that
boundary effects due to the periodicity become negligible.

Our aim is to  study the one-shot classical capacity for the
channel (\ref{finalOUT}) and analyse how it behaves with respect to the
correlation parameter $x$. By definition, it is given by
\cite{HSW}:
\begin{equation}
C=\sup_{\{p_i,\rho_i\}}\chi(\Phi(\rho_i,p_i)),
\end{equation}
where the supremum is taken over all possible ensemble of input
states, namely$\{p_i,\rho_i\}$ and $\chi$ is the Holevo
information
\begin{equation}\label{HolevoQ}
\chi(\rho)=S\left(\Phi\left(\sum_ip_i\rho_i\right)\right)-\sum_{i}p_i
S\left(\Phi\left(\rho_i\right)\right),
\end{equation}
with $S$ the von-Neumann entropy.

In the following we are going to consider classical information
over a continuous alphabet encoded in weak coherent states as
well as classical information over a discrete alphabet encoded
into qubit states.


\section{Weak coherent input}\label{wcInput}

Since coherent states turns out to be optimal for achieving 
 the capacity in lossy bosonic
channels \cite{GioPRL92}, we consider them as input to the channel (\ref{finalOUT}). 
A coherent state of complex amplitude $\alpha$ is expressed in the
Fock basis by
\begin{equation}\label{input}
|\alpha\ra=e^{-\frac{|\alpha|^2}{2}}\sum_{n=0}^{\infty}\frac{\alpha^n}{\sqrt{n!}}|n\ra.
\end{equation}
We are here interested on weak (attenuated) coherent states which
likely contain no photon or one photon, i.e. they are
characterized by an amplitude $|\alpha|\ll 1$. To provide the two
input modes of the channel in figure (\ref{channel}), we
consider  horizontal and vertical-polarized photons so that the
state of modes $1$ and $2$ in figure \ref{channel} can be written
as
\begin{equation}\label{weakinput}
|\alpha\ra\approx
N(|00\ra_{1,2}+\alpha(|01_{_V}\ra_{1,2}+|1_{_H}0\ra_{1,2}))
\end{equation}
where $N=\frac{1}{\sqrt{1+2|\alpha|^2}}$ represents the
normalization factor.

It is straightforward to show that the state (\ref{weakinput})
will be transformed to the following state after the action of
the first beam splitter:
\begin{eqnarray}
&&N[|00\ra+\alpha\cos\theta(|01_{_V}\ra+|1_{_H}0\ra)-\alpha\sin\theta(|01_{_H}\ra-|1_{_V}
0\ra)]_{3,4}\cr\cr
&&\quad=\cos\theta|\alpha\ra_{3,4}-N[\cos\theta|00\ra+\alpha\sin\theta(|01_{_H}\ra-|1_{_V}
0\ra)]_{3,4}. \label{firststep}
\end{eqnarray}
In the second step, the above state (\ref{firststep}) interacts
with modes $5$ and $6$ which are initially in the vacuum.
Hence, we have
\begin{eqnarray}
N[|0000\ra&+&\alpha\cos\theta\cos\phi(|01_{_V}00\ra+|1_{_H}000\ra)\cr
&-&\alpha\cos\theta\sin\phi(|0001_{_V}\ra+|001_{_H}0\ra)\cr
&-&\alpha\sin\theta\cos\phi(|01_{_H}00\ra-|1_{_V}000\ra)\cr
&+&\alpha\sin\theta\sin\phi(|0001_{_H}\ra-|001_{_V}0\ra)]_{7,8,9,10}.
\label{secondstep}
\end{eqnarray}
By introducing \ba
|f(\alpha)\ra&:=&|00\ra+\alpha|01_{_H}\ra-\alpha|1_{_V} 0\ra,\\
g(\theta,\phi)&:=&1-(\cos\theta-\sin\theta)(\cos\phi-\sin\phi), \ea
 we can rewrite the state (\ref{secondstep}) in the following form:
\begin{eqnarray}
|\Psi\ra=N&[&g(\theta,\phi)|0000\ra_{7,8,9,10}+\cos\theta\cos\phi|\alpha\ra_{7,8}|00\ra_{9,10}-\cos\theta\sin\phi|00\ra_{7,8}|\alpha\ra_{9,10}\cr\cr
&-&\sin\theta\cos\phi|f(\alpha)\ra_{7,8}|00\ra_{9,10}+\sin\theta\sin\phi|00\ra_{7,8}|f(\alpha)\ra_{9,10}],
\end{eqnarray}
which shows that with some probability the input state
$|\alpha\ra$ emerges at the output modes $7$ and $8$. However,
there is a chance of loosing all the photons to the environment,
and a chance that, though the photons do not go to the
environment, the output state $|f(\alpha)\ra$ is different from
the input one.

To find the final state which represents the
channel output we trace $|\Psi\ra\la\Psi|$ over modes $9$ and
$10$ and then integrate over $\theta$ and $\phi$ as in
(\ref{finalOUT}), thus obtaining
\begin{equation}\label{foutC}
\Phi\left(|\alpha\ra\la\alpha|\right)=N^2\left(\begin{array}{ccccc}
1+2|\alpha|^2(1-\mathcal{A}-\mathcal{B})&\bar{\alpha}\mathcal{D}&\bar{\alpha}\mathcal{D}&-\bar{\alpha}\mathcal{E}&-\bar{\alpha}\mathcal{E}\cr
\alpha\mathcal{D}&|\alpha|^2\mathcal{A}&|\alpha|^2\mathcal{A}&-\frac{1}{2}|\alpha|^2\mathcal{C}&-\frac{1}{2}|\alpha|^2\mathcal{C}\cr
\alpha\mathcal{D}&|\alpha|^2\mathcal{A}&|\alpha|^2\mathcal{A}&-\frac{1}{2}|\alpha|^2\mathcal{C}&-\frac{1}{2}|\alpha|^2\mathcal{C}\cr
-\alpha\mathcal{E}&-\frac{1}{2}|\alpha|^2\mathcal{C}&-\frac{1}{2}|\alpha|^2\mathcal{C}&|\alpha|^2\mathcal{B}&|\alpha|^2\mathcal{B}\cr
-\alpha\mathcal{E}&-\frac{1}{2}|\alpha|^2\mathcal{C}&-\frac{1}{2}|\alpha|^2\mathcal{C}&|\alpha|^2\mathcal{B}&|\alpha|^2\mathcal{B}\cr
\end{array}\right),
\end{equation}
where $\mathcal{A}, \mathcal{B}, \mathcal{C}, \mathcal{D}$ and
$\mathcal{E}$ are real functions of $\theta^*, \phi^*, \sigma$ and $x$
whose explicit form is given in Appendix. It is important to note
that the order of $|\alpha|$ is the same in the output state and
the input state $|\alpha\ra\la\alpha|$.

Once we have the output state of the channel we can then evaluate
the Holevo information. We recall that the Holevo information for
an ensemble $\mathfrak{S}=\{|\alpha\ra;\;\wp(\alpha)\}$ of input
states corresponding to a continuous alphabet is given by:
\begin{equation}\label{HolevoC}
\chi(\mathfrak{S})=S(\tilde{\rho})-\int d^2\alpha
\wp(\alpha)S\left(\Phi(|\alpha\ra\la\alpha|)\right),
\end{equation}
in which $\wp$ is a probability distribution and $\tilde{\rho}$ is
the average output state
\begin{equation}
\tilde{\rho}=\Phi\left(\int d^2\alpha
\wp(\alpha)|\alpha\ra\la\alpha|\right)=\int d^2\alpha
\wp(\alpha)\Phi\left(|\alpha\ra\la\alpha|\right).
\end{equation}
We restrict the ensemble of input states to that of weak coherent
states (\ref{weakinput}) with Gaussian distribution $\wp(\alpha)$:
\begin{equation}\label{Censemble}
\wp(\alpha)=\frac{1}{2\pi\Delta^2}e^{-\frac{|\alpha|^2}{2\Delta^2}}.
\end{equation}
The variance $\Delta^2$ should be chosen in a way that the
ensemble of input states only includes weak coherent states, i.e.
$\Delta\ll 1$. We also know that $\theta$ and
$\phi$ are random variables with Gaussian probability
distribution centered around $\theta^*$ and $\phi^*$ with
variance $\sigma$. Therefore to have the channel close to an
ideal transmitter, we assume narrow probability distributions
(small $\sigma$) centered around $\theta^*=\phi^*=0$. Under these
assumptions $\mathcal{C}=\mathcal{E}\approx 0$, therefore the
output of the channel (\ref{foutC}) is block diagonal with the
following eigenvalues:
\begin{eqnarray}
\lambda_{1,2}&=&\frac{N^2}{2}(1+2|\alpha|^2-2|\alpha|^2\mathcal{B}\pm\sqrt{(1-2|\alpha|^2\mathcal{A})^2+8|\alpha|^2\mathcal{D}},\cr
\lambda_{3}&=&2N^2|\alpha|^2\mathcal{B},\cr \lambda_{4,5}&=&0.
\end{eqnarray}
Using these eigenvalues the output entropy
\begin{equation}
S\left(\Phi(|\alpha\ra\la\alpha|)\right)=-\sum_{i=1}^5
\lambda_i\log\lambda_i,
\end{equation}
can be easily expressed analytically. Nevertheless, the integral
in the second term of the Holevo information (\ref{HolevoC})
should be computed numerically.

\begin{figure}[t]
 \centering
  \includegraphics[width=12cm,height=8.5cm,angle=0]{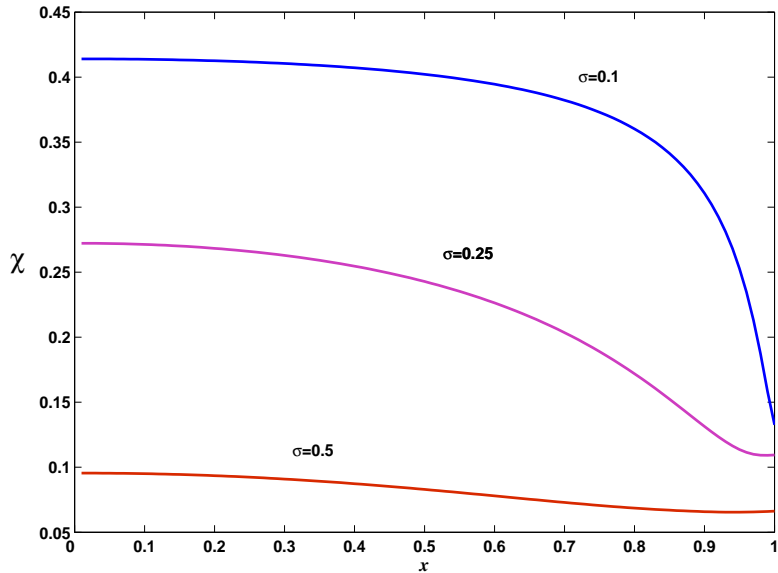}
 \caption{Holevo information versus correlation parameter $x$ when information is encoded in weak coherent states. Different lines are for different values
  of $\sigma$.}
\label{Cholevo}
\end{figure}

For finding the first term of Holevo information we note that
$\wp(\alpha)$ is even in $\alpha$, therefore the average output
state is block diagonal
\begin{equation}
\tilde{\rho}=\left(\begin{array}{ccccc} 1-
2\mathcal{L}(\mathcal{A}+\mathcal{B})& &
 &  &  \cr
 &\mathcal{A}\mathcal{L}&\mathcal{A}\mathcal{L}&  &  \cr
 &\mathcal{A}\mathcal{L}&\mathcal{A}\mathcal{L}&  &  \cr
 &  &  &\mathcal{B}\mathcal{L}&\mathcal{B}\mathcal{L}\cr
 &  &  &\mathcal{B}\mathcal{L}&\mathcal{B}\mathcal{L}
\end{array}\right),
\end{equation}
with the following eigenvalues:
\begin{eqnarray}
\tilde{\lambda}_1&=&1+2\mathcal{L}(1-\mathcal{A}-\mathcal{B}),\cr
\tilde{\lambda}_2&=&2\mathcal{A}\mathcal{L},\cr
\tilde{\lambda}_3&=&2\mathcal{B}\mathcal{L},\cr
\tilde{\lambda}_{4,5}&=&0,
\end{eqnarray}
and
\begin{equation}
\mathcal{L}=\int d^2\alpha
\wp(\alpha)\frac{|\alpha|^2}{1+2|\alpha|^2}.
\end{equation}
Evaluating $\mathcal{L}$ numerically the first term of Holevo
information in equation (\ref{HolevoC})
\begin{equation}
S(\tilde{\rho})=-\sum_i\tilde{\lambda}_i\log \tilde{\lambda}_i,
\end{equation}
can also be found. The final result for Holevo information
(\ref{HolevoC}) is plotted in figure (\ref{Cholevo}) versus the
correlation parameter $x$, for different values of $\sigma$. It
shows that the Holevo information for the ensemble of input
states (\ref{Censemble}) is a decreasing function of correlation
parameter $x$. For $x=0$ polarization mixing and loss happen
independently, therefore there is the chance that just one of
them happens. By increasing the correlation parameter the
probability that both errors happen increases and therefore the
rate of information transmission decreases. As it is clear from
the figure (\ref{Cholevo}), by decreasing $\sigma$ the Holevo
quantity increases, this reflects the fact the narrow the
probability distribution is, the less is the noise in  the
channel. As $\sigma$ goes to $0$,
$\mathcal{A},\mathcal{D}\rightarrow1$ and
$\mathcal{B}\rightarrow0$. In this case the output state gets
closer to a pure state of the form of input state
(\ref{weakinput}).

\section{Qubit inputs}\label{pInput}

In this section we use qubit encoding in single photons. To this
end we consider the two input modes 1 and 2 to only have one
photon in total. Therefore the photon is either in mode 1 or in
mode 2 and a general input pure state is given by:
\begin{equation}\label{DRstate}
|\psi\ra=c_0|01\ra+c_1|10\ra,
\end{equation}
with $c_0,c_1$ are complex coefficient such that
$|c_0|^2+|c_1|^2=1$. By referring to the dual-rail coding
\begin{equation}
|0\ra_L\leftrightarrow |01\ra_{1,2} \hskip 1cm
|1\ra_L\leftrightarrow |10\ra_{1,2}.
\end{equation}
it is clear that the state in (\ref{DRstate}) represent a general
pure state for a single logical qubit.

Our aim is to find the capacity of channel (\ref{finalOUT}) when
we encode information in qubits. The upper bound of Holevo
information (\ref{HolevoQ}) is given by
\begin{equation}
\chi\leq 1-S(\Phi(\rho^*)),
\end{equation}
in which $\rho^*$ is an input state with minimum output entropy
\cite{M3}. Using the concavity property of entropy, it has been
shown that for finding the input state with minimum output
entropy it is enough to search among the pure states \cite{M3}.
Therefore we consider the input ensemble that includes pure states
of form (\ref{DRstate}). Using equation (\ref{U}) it is easy to
show that
\begin{eqnarray}
U(\theta)|01\ra&=&\cos\theta|01\ra-\sin\theta|10\ra,\cr
U(\theta)|10\ra&=&\sin\theta|01\ra+\cos\theta|10\ra,
\end{eqnarray}
therefore the action of the first beam splitter is a rotation
by angle $\theta$ in the $|0\ra_L$ and $|1\ra_L $ plain, and the
input state
\begin{equation}
|\psi\ra_L=c_0|0\ra_L+c_1|1\ra_L,
\end{equation}
will be transformed to
\begin{equation}
|\psi(\theta)\ra_L=e^{-i\theta\sigma_y}|\psi\ra_L,
\end{equation}
in which $\sigma_y$ is the $y$ Pauli operator. The dual-rail
state at the output modes of the first beam splitter is then
given by
\begin{equation}
(c_0\cos\theta+c_1\sin\theta)|01\ra_{3,4}-(c_0\sin\theta-c_1\cos\theta)|10\ra_{3,4}.
\end{equation}
This state interacts with modes $5$ and $6$ which are in the
vacuum and will be transformed into
\begin{eqnarray}
&&(c_0\cos\theta+c_1\sin\theta)[\sin\phi|0001\ra+\cos\phi|0100\ra]_{7,8,9,10}\cr
&-&(c_0\sin\theta-c_1\cos\theta)[\sin\phi|0010\ra+\cos\phi|1000\ra]_{7,8,9,10}.
\end{eqnarray}
If we take $|2\ra_L\leftrightarrow |00\ra$, then the output state
of system and environment can be summarized as follows:
\begin{equation}
\sin\phi|2_s\ra_L|\psi(\theta)_e\ra_L+\cos\phi|\psi(\theta)_s\ra_L|2_e\ra_L,
\end{equation}
where $s$ and $e$ are labels for system and environment
respectively. Tracing over the environment and taking into
account the correlation between the parameters of the beam
splitters the final output state of the channel is:
\begin{equation}\label{Qfinal}
\Phi(|\psi\ra_L\la\psi|)=\epsilon|2\ra_L\la
2|+(1-\epsilon)e^{-i\theta^*\sigma_y}|\psi\ra_L\la\psi|e^{i\theta^*\sigma_y},
\end{equation}
where
\begin{equation}
\epsilon=\int \sin^2\phi P(\theta,\phi) d\theta d\phi
=\frac{1}{2}\left(1-\cos
2\phi^*e^{-\frac{2\sigma^2}{1-x^2}}\right).
\end{equation}
Since $|2\ra_L$ is orthogonal to the manifold of the input states,
equation (\ref{Qfinal}) shows that the information is lost with
probability $\epsilon$ and the state is rotated by average angle $\theta^*$ 
with probability $(1-\epsilon)$. Therefore the total map is a
kind of a quantum  erasure channel \cite{erasure} and its
capacity is given by
\begin{equation}
C=(1-\epsilon)C_{\Omega},
\end{equation}
where $C_{\Omega}$ is the capacity of the the channel $\Omega$
\begin{equation}
\Omega(\rho)=e^{-i\theta^*\sigma_y}\rho e^{i\theta^*\sigma_y}.
\end{equation}
\begin{figure}[t]
 \centering
  \includegraphics[width=12cm,height=8.5cm,angle=0]{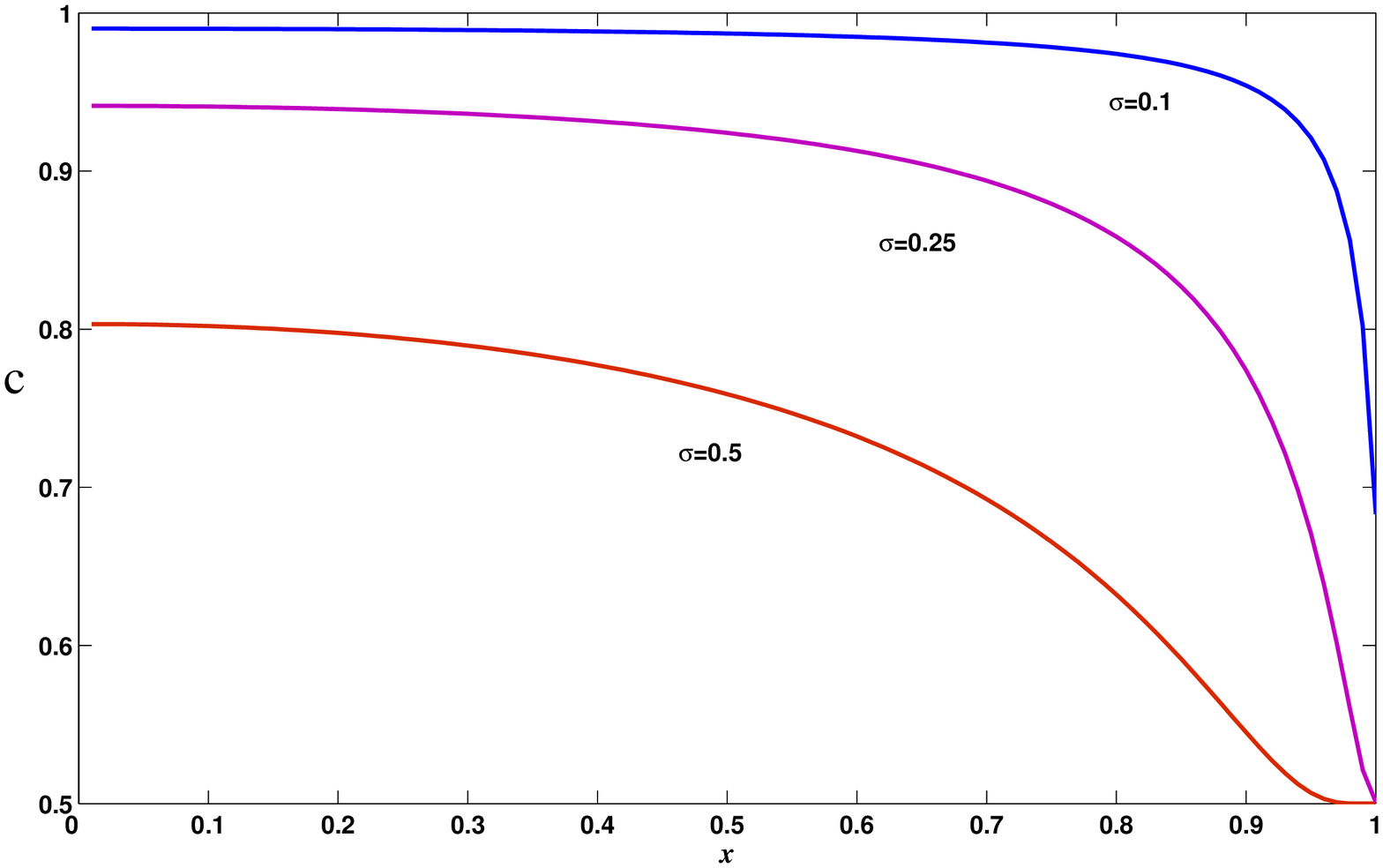}
 \caption{Classical capacity of the channel versus correlation parameter $x$ when information is encoded into qubits. Different lines are for different values of $\sigma$.}
\label{figerasure}
\end{figure}
Since $\Omega$ is a rotation about $y$ axis it is obvious that
the following states remain invariant under its action:
\begin{equation}
|\psi_{1,2}\ra_L=\frac{1}{\sqrt{2}}(|0\ra_L\pm i|1\ra_L),
\end{equation}
thus giving zero output entropy and minimizing the second term of
Holevo information.

It is easy to show that if these states are used with equal
probability in the ensemble of input state then
\begin{equation}
\frac{1}{2} \Omega (|\psi_1\ra_L\la\psi_1|) +\frac{1}{2} \Omega
(|\psi_2\ra_L\la\psi_2|)=\frac{1}{2}I.
\end{equation}
Therefore this ensemble also maximizes the first term in Holevo
information and $C_{\Omega}=1$ can be achieved by using the above
states to encode information. As a consequence the capacity of the
channel $\Phi$ results $C=(1-\epsilon)$.

Figure (\ref{figerasure}) shows this capacity versus the
correlation parameter $x$ for different values of $\sigma$ when
$\theta^*=\phi^*=0$. At $x=0$ the values of $\phi$ which are close
to $0$ are the most probable values, however by increasing $x$ the
higher values of $\phi$ also have chance to happen. Therefore the
probability of loosing information increases and as it is seen in
figure (\ref{figerasure}) the capacity of the channel decreases.
By decreasing $\sigma$ the probability distribution becomes more
and more narrow and the chance for having values of $\phi$ far
from $\phi^*=0$ become less. Therefore the channel capacity
increases by decreasing the variance of probability distribution
which can be seen in figure (\ref{figerasure}).


\section{conclusion}\label{conclusion}

We have provided a model of quantum channel with correlated
errors of different kinds. While correlation in errors between
different uses of channels have been studied before, in our model
different types of correlated errors happen in each use of the channels
on different degrees of freedom. This assumption can be justified by considering practical situations
i. e. optical fibers where photon loss and polarization errors may be correlated. Actually our model describes photon losses
depending on the degree of polarization mixedness.

On the one hand, for classical  information over a continuous
alphabet we have considered weak coherent states and we have
shown that the Holevo information decreases by increasing the
correlation between polarization mixdness and losses. On the
other hand,  for classical information over a discrete alphabet
we have considered single photons using dual rail representation.
In this case we have shown that the channel is kind of quantum
erasure channel where the probability of erasing information
increases with the polarization mixdness. It is worth remarking
that while often memory effects lead to an enhancement of the
channel capacity, in the present model exactly the opposite happens. 
This is mainly due to the fact that the polarization
mixing alone cannot be considered as a ``true" error, i.e. it is
not due to the action of an environment.

In conclusion, we are confident that this work may pave the way for characterizing 
realistic quantum channels where different
degrees of freedom and different effect are involved.


\acknowledgements

L. M thanks C. Cafaro, R. Kumar and C. Lupo for valuable
discussions. This work is supported by European Commission under
the FET-Open grant agreement CORNER, number FP7-ICT-213681.


\section*{Appendix}

To evaluate the quantities $\mathcal{A}$, $\mathcal{B}$,
$\mathcal{C}$, $\mathcal{D}$ and $\mathcal{E}$ in (\ref{foutC})
we use the fact that the variance of the probability distribution
$P(\theta,\phi)$ is sufficiently smaller than $2\pi$, hence we
can extend the limits of the integrals to $(-\infty,\infty)$
treating them like standard Gaussian integrals, thus obtaining:
\begin{eqnarray}
\mathcal{A}&=&\int \cos^2\theta\cos^2\phi P(\theta,\phi)d\theta
d\phi\cr&=&\frac{1}{4}\left(1+e^{-\frac{2\sigma^2}{1-x^2}}(\cos
2\theta^*+\cos
2\phi^*)+\frac{1}{2}(e^{-\frac{4\sigma^2}{1-x}}+e^{-\frac{4\sigma^2}{1+x}})\cos
2\theta^*\cos 2\phi^*\right),\cr\cr
\mathcal{B}&=&\int \sin^2\theta \cos^2\phi P(\theta,\phi)d\theta
d\phi\cr&=&\frac{1}{4}\left(1-e^{-\frac{2\sigma^2}{1-x^2}}(\cos
2\theta^*-\cos
2\phi^*)-\frac{1}{2}(e^{-\frac{4\sigma^2}{1-x}}+e^{-\frac{4\sigma^2}{1+x}})\cos
2\theta^*\cos 2\phi^*\right),\cr\cr
\mathcal{C}&=&\int \sin 2\theta\cos^2\phi P(\theta,\phi)d\theta
d\phi=\frac{1}{4}\left(2e^{-\frac{2\sigma^2}{1-x^2}}-(e^{-\frac{4\sigma^2}{1-x}}+e^{-\frac{4\sigma^2}{1+x}})\cos\phi^*\right)\sin
2\theta^*,\cr\cr
\mathcal{D}&=&\int \cos\theta\cos\phi P(\theta,\phi)d\theta
d\phi=\frac{1}{2}\left(e^{-\frac{\sigma^2}{1-x}}+e^{-\frac{\sigma^2}{1+x}}\right)\cos\theta^*\cos\phi^*,\cr\cr
\mathcal{E}&=&\int \sin\theta\cos\phi P(\theta,\phi)d\theta
d\phi=\frac{1}{2}\left(e^{-\frac{\sigma^2}{1-x}}+e^{-\frac{\sigma^2}{1+x}}\right)\sin\theta^*\cos\phi^*.
\end{eqnarray}


\end{document}